\pdfoutput=1
\documentclass[11pt]{article}
\usepackage{acl}
\usepackage{times}
\usepackage{latexsym}
\usepackage[T1]{fontenc}
\usepackage[utf8]{inputenc}
\usepackage{microtype}
\usepackage{inconsolata}
\usepackage{enumitem}
\usepackage{graphicx}
\usepackage{multirow}
\usepackage{amsmath}
\usepackage{amssymb}
\usepackage{algorithm}
\usepackage{algorithmicx}
\usepackage{algpseudocode}
\usepackage{xcolor}
\usepackage{array}
\usepackage{booktabs}
\usepackage{arydshln}

\title{Generative Dense Retrieval: Memory Can Be a Burden}
\author{Peiwen Yuan$^1$\footnotemark[1], Xinglin Wang$^1$\footnotemark[1], Shaoxiong Feng$^2$, Boyuan Pan$^2$, Yiwei Li$^1$\\ {\bf Heda Wang$^2$, Xupeng Miao$^3$, Kan Li$^{1}$\footnotemark[2]}\\
  $^1$School of Computer Science and Technology, Beijing Institute of Technology \\
  $^2$Xiaohongshu Inc \quad
  $^3$Carnegie Mellon University\\
  \texttt{\{peiwenyuan,wangxinglin,liyiwei,likan\}@bit.edu.cn} \ \ \texttt{\{xupeng\}@cmu.edu}\\
  \texttt{\{shaoxiongfeng2023,whd.thu\}@gmail.com} \ \  \texttt{\{panboyuan\}@xiaohongshu.com}}

\begin{document}
\maketitle
\renewcommand{\thefootnote}{\fnsymbol{footnote}} 
\footnotetext[1]{Equal contribution.} 
\footnotetext[2]{Corresponding author.} 

\renewcommand{\thefootnote}{\arabic{footnote}}
\begin{abstract}
Generative Retrieval (GR), autoregressively decoding relevant document identifiers given a query, has been shown to perform well under the setting of small-scale corpora.
By memorizing the document corpus with model parameters, GR implicitly achieves deep interaction between query and document.
However, such a memorizing mechanism faces three drawbacks:
(1) Poor memory accuracy for fine-grained features of documents;
(2) Memory confusion gets worse as the corpus size increases;
(3) Huge memory update costs for new documents.
To alleviate these problems, we propose the \textbf{G}enerative \textbf{D}ense \textbf{R}etrieval (GDR) paradigm. 
Specifically, GDR first uses the limited memory volume to achieve inter-cluster matching from query to relevant document clusters. 
Memorizing-free matching mechanism from Dense Retrieval (DR) is then introduced to conduct fine-grained intra-cluster matching from clusters to relevant documents. 
The coarse-to-fine process maximizes the advantages of GR's deep interaction and DR's scalability. 
Besides, we design a cluster identifier constructing strategy to facilitate corpus memory and a cluster-adaptive negative sampling strategy to enhance the intra-cluster mapping ability.
Empirical results show that GDR obtains an average of 3.0 $\mathrm{R}@100$ improvement on NQ dataset under multiple settings and has better scalability\footnote{Our code have been released on \url{https://github.com/ypw0102/GDR}.}.

\end{abstract}

\section{Introduction}
Text retrieval \citep{DBLP:conf/emnlp/KarpukhinOMLWEC20,DBLP:journals/corr/abs-2211-14876} is an essential stage for search engines \citep{DBLP:conf/www/BrickleyBN19}, question-answering systems \citep{DBLP:conf/acl/LiuGFYCJLD20} and dialog systems \citep{DBLP:journals/sigkdd/ChenLYT17}. 
Traditional retrieval methods include \textit{sparse retrieval} (SR) and \textit{dense retrieval} (DR). 
SR \citep{DBLP:journals/ftir/RobertsonZ09,DBLP:conf/sigir/RobertsonW97} relies on the assumption that queries and relevant documents have a high degree of word overlap.
However, such methods suffer from the zero-recall phenomenon when there is a lexical mismatch between queries and documents.
% It is possible to recall the most relevant documents when the query contains accurate keywords. However, when there is a lexical mismatch between query and document, it is likely that no relevant document can be hit. 
% However, semantically relevant documents can not be hit once there is a lexical mismatch between queries and documents.
DR \citep{DBLP:conf/emnlp/RenQLZSWWW21,DBLP:conf/iclr/ZhangGS0DC22} alleviates this issue by training dual-encoders for semantic matching instead of lexical matching, which brings a high hit rate. 
% Nevertheless, the differences in semantics among multiple relevant documents for a given query may make them not close to each other in the semantic space, making it challenging for a single query representation to recall all the relevant documents simultaneously \citep{DBLP:conf/acl/ZhangLGJD22}.
% Nevertheless, most queries are semantically related to multiple documents that may not be close to each other in the semantic space. Thus it is challenging for DR to use a single query representation to recall all the relevant documents \citep{DBLP:conf/acl/ZhangLGJD22}. 
Nevertheless, most queries are semantically related to multiple documents that may not be close to each other in semantic space. Thus it is challenging to use a single query representation to recall all the relevant documents with matching mechanism \citep{DBLP:conf/acl/ZhangLGJD22}.

Recently, \textit{generative retrieval} (GR) \citep{DBLP:journals/corr/abs-2208-09257,DBLP:conf/nips/BevilacquaOLY0P22}, which utilizes a language model to memorize document features and autoregressively decodes the identifiers of relevant documents given a query, is considered a promising paradigm. 
The model is served as a memory bank for candidate documents, and the memorizing process implicitly implements the deep interaction between queries and documents by attention mechanism, which has been proven to be effective in the small-scale corpus settings \citep{DBLP:conf/nips/WangHWMWCXCZL0022,DBLP:journals/corr/abs-2304-04171}. 
Also, beam search, a diversity-promoting decoding strategy, is beneficial for the model to find relevant documents from multiple directions and thus can recall more relevant documents than DR \citep{DBLP:conf/nips/Tay00NBM000GSCM22}.

However, after empirically comparing the performance of typical GR model NCI \citep{DBLP:conf/nips/WangHWMWCXCZL0022} and DR model AR2 \citep{DBLP:conf/iclr/ZhangGS0DC22}, we found that the memorizing mechanism brings three problems:
(1) \textit{Poor memory accuracy for fine-grained features of documents}.
We calculated the error rate of each position when decoding document identifiers (see Table \ref{tab: prediction error rate}). Compared with AR2, NCI performs well on the former part of the decoding process while poorly on the latter part. 
We argue that NCI aims to map queries to relevant document identifiers instead of real document content, which results in its lack of accurate memory for fine-grained document features. 
(2) \textit{Memory confusion gets worse as the corpus size increases}.
As shown in Table \ref{tab: NCI scale}, we scaled both training and candidate corpus sizes from 334K to 1M and found that NCI decreased by 11.0 on $\mathrm{R}@100$ while AR2 only decreased by 2.8. 
NCI trained on 1M training corpus is further tested on 334K candidate corpus. The results indicate that the burden of memorizing more documents causes 5.7 $\mathrm{R}@100$ drop. % It verifies that memory confusion will aggravate as the training corpus size grows.
% We further evaluated NCI trained on 1M corpus (required to memorize 1M documents) on 334K corpus (row 2). The results indicate that from 334K to 1M, the burden of memorizing more documents accounts for 5.7 R@100 drop while the addition of new hard negatives accounts for the other 5.3 R@100 drop. 
% This proves that the inaccurate documents memory caused by the limited memory volume will lead to significant performance degradation under large-scale corpus setting.
(3) \textit{Huge memory update costs for new documents}.
When new documents come, the document cluster tree needs to be updated, and the model needs to be re-trained to re-memorize all the documents. 
Otherwise, the outdated mapping relationship, i.e., query to document identifiers and document identifiers to documents, will significantly degrade the retrieval performance (see Table \ref{tab: scalability of model}).

Based on the above analysis, a natural idea is to employ memorizing-free matching mechanism from DR to alleviate the burden faced by the memorizing mechanism. However, it is challenging to realize complementary advantages of both mechanisms while ensuring retrieval efficiency. To this end, we propose a coarse-to-fine retrieval paradigm \textbf{G}enerative \textbf{D}ense \textbf{R}etrieval (GDR). Concretely, memorizing mechanism and matching mechanism are successively applied to achieve coarse-grained inter-cluster (query $\rightarrow$ document clusters) and fine-grained intra-cluster (document clusters $\rightarrow$ documents) matching. A shared query encoder is used to generate query representations that apply both mechanisms, thereby improving retrieval efficiency. We also explore the strategy of constructing a memory-friendly document cluster tree, including distinguishable document clusters and controllable cluster amounts, so as to further alleviate memory burden.
Moreover, a cluster-adaptive negative sampling strategy is proposed to enhance the intra-cluster matching ability of GDR.

\begin{table}[t]
    \small
    \centering
    % The i$^{th}$ decoding result is counted as wrong when 
    % For the top-k docids recalled by the model, We estimate the error rate by calculating the probability that, given the prefix of the recalled doc identifier up to the i-$1^{th}$ position belonging to a corresponding label doc identifier, the addition of the i$^{th}$ position prediction does not belong to any prefix of a label doc identifier.}
    \begin{tabular}{lcccccc}
    \hline
    \multirow{2}{*}{Model} & \multicolumn{6}{c}{Error Rate of the i$^{th}$ Position} \\ % identifier prediction
    \cline{2-7}
     &1$^{st}$ & 2$^{nd}$ & 3$^{rd}$ &  4$^{th}$  & 5$^{th}$ & 6$^{th}$ \\
    \hline
    NCI	& \textbf{1.09} & \textbf{1.75}	& \textbf{1.86}	& 5.77	& 14.91 & 12.66\\
    AR2	& 1.19	& 1.77	& 2.11	& \textbf{5.44}	& \textbf{8.03}	& \textbf{3.05}\\
    \hline
    \end{tabular}
    \caption{Error rate (\%) on the i$^{th}$ position when decoding document identifiers. See Appendix \ref{sec:calculation of error rate} for the detailed calculation method.}
    \label{tab: prediction error rate}
    \footnotesize{}
\end{table}

\definecolor{lightgreen}{RGB}{48,128, 20} % 自定义绿色

\begin{table}[t]
    \small
    \centering
    % : scale of training corpus - scale of candidate documents during the evaluating phrase .} 
    %For the setting x-y, the values of x and y represent the number of corpora in the training set and the number of candidate documents during the evaluating phrase respectively.}
    \setlength{\tabcolsep}{0.34em} % 设置列之间的间距
    \begin{tabular}{lcc}
    \hline
    \multirow{2}{*}{Settings} & NCI\hspace{0.25cm} & AR2\hspace{0.25cm} \\
    \cline{2-3}
     &R@1/100\hspace{0.25cm} &R@1/100\hspace{0.25cm} \\
    \hline
    334K-334K  & 14.7\hspace{0.25cm}-\hspace{0.25cm} / 65.5\hspace{0.25cm}-\hspace{0.25cm} \  & 21.2\hspace{0.25cm}-\hspace{0.2cm} / 69.0\hspace{0.25cm}-\hspace{0.2cm}\  \\
    1M-1M & 11.1\textcolor{lightgreen}{$\downarrow$3.6} / 54.5\textcolor{lightgreen}{$\downarrow$11.0} & 20.3\textcolor{lightgreen}{$\downarrow$0.9} / 66.2\textcolor{lightgreen}{$\downarrow$2.8} \\ 
    1M-334K &  12.3\textcolor{lightgreen}{$\downarrow$2.4} / 59.8\textcolor{lightgreen}{$\downarrow$5.7}\ \ \  & 21.2\hspace{0.25cm}-\hspace{0.2cm} / 69.0\hspace{0.25cm}-\hspace{0.2cm}\  \\
    \hline
    \end{tabular}
    \caption{Performance of NCI and AR2 on NQ validation set with different settings. For setting $x-y$, $x$ denotes the training corpus size and $y$ denotes the candidate corpus size during the inference phase. AR2 is only trained on the training set, thus is independent of $x$.}
    \label{tab: NCI scale}
\end{table}

Overall, the coarse-to-fine process maintains the advantages of the memorizing mechanism while alleviating its drawbacks by introducing matching mechanism. 
Unlike GR, the limited memory volume of GDR is only responsible for memorizing the coarse-grained features of corpora. % and achieves deep interactions between query and relevant clusters. 
The fine-grained features of documents are extracted into dense representations, which promotes accurate intra-cluster mapping. 
When new documents come, GDR achieves scalability by adding documents to relevant clusters and extracting their dense representations by a document encoder, without reconstructing document identifiers and retraining the model.

Our contributions are summarized as follows:
\begin{itemize}[leftmargin=15pt]
\setlength{\itemsep}{0pt}
\setlength{\parsep}{0pt}
\setlength{\parskip}{0pt}
\item We revisit generative retrieval (GR) with a detailed empirical study, and discuss three key drawbacks that limit GR performance. % We discussed three issues of GR models, and verified them through detailed experiments.
\item We propose generative dense retrieval (GDR), a coarse-to-fine retrieval paradigm, that exploits the limited memory volume more appropriately, enhances fine-grained feature memory, and improves model scalability. % which alleviates the memory volume limitation, inaccurate fine-grained features memory, and poor scalability of GR by introducing dense feature matching and reallocating retrieval sub-tasks reasonably.
% \item We design two strategies to make document cluster tree memory-friendly and discriminate the negative documents more accurately, respectively. 
\item Comprehensive experiments demonstrate that GDR obtains higher recall scores than advanced SR, DR and GR methods. And the scalability of GDR is also significantly improved. % the excellent retrieval ability and favorable scalability of GDR.
\end{itemize}

\section{Related Work}

Given queries, text retrieval task aims to find relevant documents from a large corpus. In this section, we introduce typical paradigms DR and GR that are most related to our work.

\subsection{Dense Retrieval}

DR 
\citep{DBLP:conf/emnlp/KarpukhinOMLWEC20,DBLP:conf/iclr/XiongXLTLBAO21,DBLP:conf/emnlp/RenQLZSWWW21,DBLP:conf/acl/ZhangLGJD22,DBLP:conf/iclr/ZhangGS0DC22,DBLP:journals/corr/abs-2211-14876} is the most widely studied retrieval paradigm in recent years. A dual-encoder architecture (query-encoder and document-encoder) is commonly used to extract the dense semantic representations of queries and documents. The similarities between them are computed through simple operations (\textit{e.g.,} inner product) in Euclidean space and ranked to recall the relevant documents. By extracting features and constructing indexes for matching, DR does not have to memorize the corpus and attains good scalability.
However, the upper bound of DR is constrained due to the limited interaction between queries and candidate documents \citep{DBLP:journals/corr/abs-2208-04232}. GDR inherits the matching mechanism from DR in the fine-grained mapping stage, and introduces deep interaction through memorizing mechanism in the coarse-grained mapping stage, thus achieving better recall performance.

% While it is true that incorporating both queries and documents into one encoder for deep interaction can bring improved retrieval performance, the significant time cost associated with this approach makes it impractical for real-world applications. 
% The advancements in pre-trained language models have contributed to the state-of-the-art performance of DR in text retrieval \citep{DBLP:conf/iclr/ZhangGS0DC22}. 

\subsection{Generative Retrieval}

Recently, a new retrieval paradigm named GR, which adopts autoregressive model to generate relevant document identifiers, has drawn increasing attention. % been widely studied. 
\citet{DBLP:conf/iclr/CaoI0P21} proposes to retrieve documents by generating titles. 
\citet{DBLP:conf/nips/Tay00NBM000GSCM22} utilizes BERT \citep{DBLP:conf/naacl/DevlinCLT19} combined with the \textit{K}-means algorithm to generate identifiers with hierarchical information. 
\citet{DBLP:conf/nips/BevilacquaOLY0P22} leverages n-grams to serve as identifiers. 
\citet{DBLP:conf/nips/WangHWMWCXCZL0022} enhances the model's memory of candidate documents through query generation. 
\citet{DBLP:journals/corr/abs-2212-09744} proposes retraining model with generated queries of old documents when new documents are added to reduce forgetting. 
\citet{DBLP:journals/corr/abs-2304-04171} suggests training the model to learn to assign document identifiers. 
However, all of these methods require models to memorize the whole corpus and inevitably face the problems we have discussed above, for which we propose GDR.
% the contradiction between the memory capacity and the requirement to remember all the documents hinders the performance of the model on larger candidate sets. Additionally, when new documents are added, the model either needs to be retrained or faces the challenge of multiple documents being assigned the same identifier, resulting in poor scalability.

%%%%%%%%%%%%%%%%%%%%%%%%%%%%%%%%%%%%%%%%%%%%%%%%%%%%%%%%%%%%%%%%%%%%%%%%%%%%%%%%%%%%%%%%%%%%%%%%%%%%%%%%%%%%%%
%%%%%%%%%%%%%%%%%%%%%%%%%%%%%%%%%%%%%%%%%%%%%%%%%%%%%%%%%%%%%%%%%%%%%%%%%%%%%%%%%%%%%%%%%%%%%%%%%%%%%%%%%%%%%%
%%%%%%%%%%%%%%%%%%%%%%%%%%%%%%%%%%%%%%%%%%%%%%%%%%%%%%%%%%%%%%%%%%%%%%%%%%%%%%%%%%%%%%%%%%%%%%%%%%%%%%%%%%%%%%
%%%%%%%%%%%%%%%%%%%%%%%%%%%%%%%%%%%%%%%%%%%%%%%%%%%%%%%%%%%%%%%%%%%%%%%%%%%%%%%%%%%%%%%%%%%%%%%%%%%%%%%%%%%%%%

\section{Methodology}

\begin{figure*}[t]
    \centering
    \includegraphics[width=1.0\linewidth]{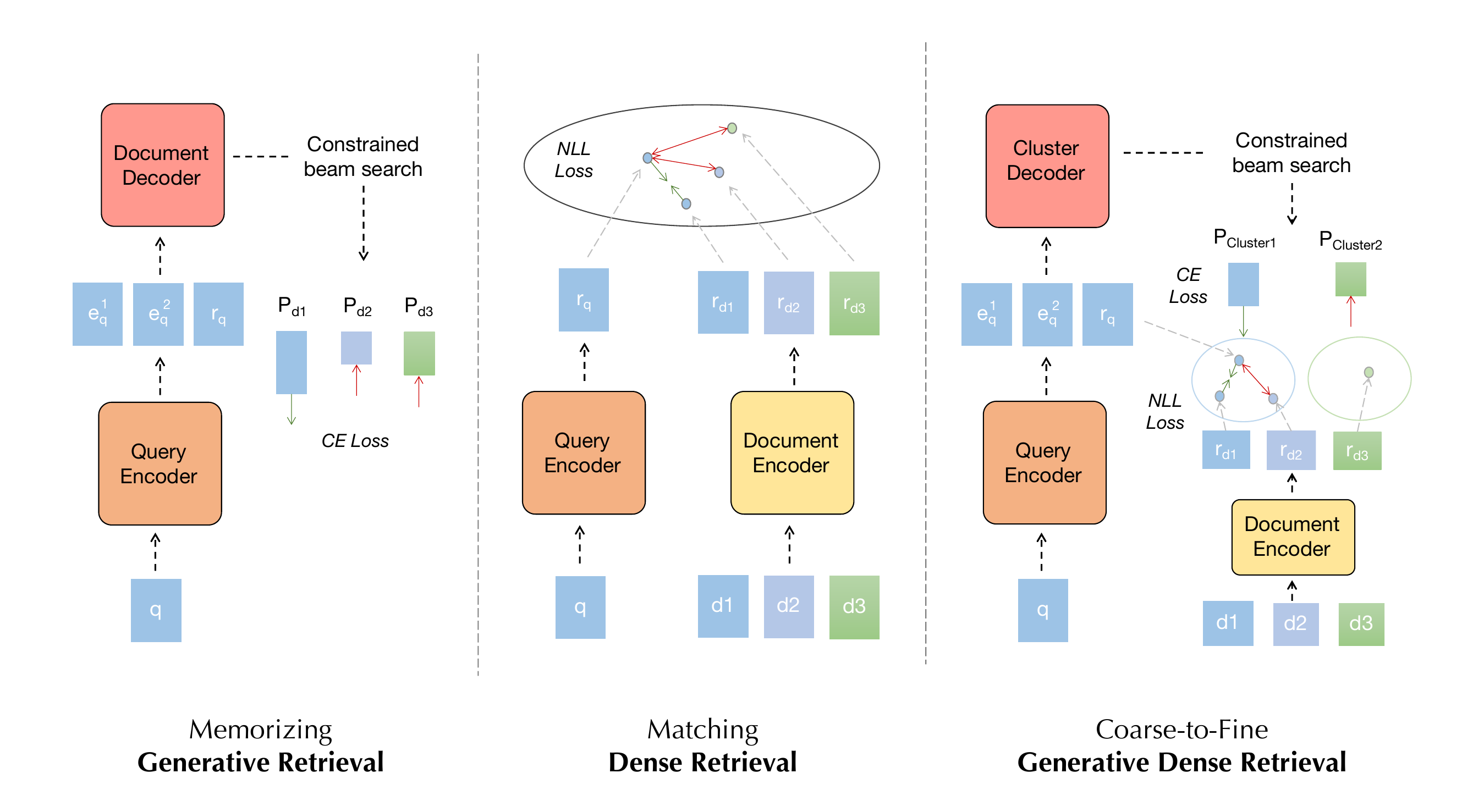}
    \caption{Illustration of Dense Retrieval, Generative Retrieval and Generative Dense Retrieval.}
    \label{fig: overview}
    % \vspace{-1.0\baselineskip}nnnj
\end{figure*}

 Our task is to retrieve a candidate document set $\mathcal{D}_c$ from a large corpus $\mathcal{D}_l$ ($|\mathcal{D}_l|>>|\mathcal{D}_c|$) for a given query $q$, with the objective of including as many documents $d$ from $\mathcal{D}_q$ as possible, where $\mathcal{D}_q$ is the set of documents relevant to $q$. In this section, we introduce the proposed Generative Dense Retrieval (GDR) paradigm (see Figure \ref{fig: overview}). To realize complementary advantages of memorizing mechanism and matching mechanism, we need to consider the following issues:

 % Basically, GDR utilizes the advantage of GR in inter-cluster coarse-grained matching to recall relevant document clusters, and then leverages the superiorities of DR in fine-grained representation extracting, better scalability and memorizing-free characteristics to further retrieve the most relevant documents from the recalled clusters.

\subsection{Order of Applying Two Mechanisms}
% 通过表1，2的实验，我们发现query到文档之间共性的粗粒度特征匹配由记忆机制实现能取得更低的error rate，而海量文档各自的细粒度特征则更适合由特征抽取与匹配机制实现。
%因此，我们考虑utilizes the advantage of memorizing mechanism in deep interaction between query and corpus memory bank to recall relevant document clusters, and then leverages the superiorities of matching mechanism in fine-grained representation extracting, better scalability and memorizing-free characteristics to further retrieve the most relevant documents from the recalled clusters.
Based on Table \ref{tab: prediction error rate} and Table \ref{tab: NCI scale}, we found that the coarse-grained semantic mapping between query and documents attained lower error rates when applying memorizing mechanism (NCI), while feature extraction and matching mechanism (AR2) was better suited for handling fine-grained features of numerous documents. 
Thus, we consider utilizing the advantage of memorizing mechanism in deep interaction between query and corpus memory bank to recall relevant document clusters. 
Afterwards, we leverage the superiorities of memorizing-free matching mechanism in fine-grained representation extracting and better scalability characteristics to further retrieve the most relevant documents from the recalled clusters.

\paragraph{Inter-cluster Matching}
The classic Encoder-Decoder architecture is used to achieve the inter-cluster mapping $f_{inter}: q \rightarrow \mathrm{CID}^{1:k}$, where $\mathrm{CID}$ denotes document cluster identifiers. Given query $q^{1:|q|}$, GDR first leverages Query Encoder $E_Q$ to encode it into query embeddings $e_q^{1:|q|} \in \mathbb{R}^d$ and takes the embedding of <$CLS$> token as query representation $r_q$. Based on this, the probability of generating $\mathrm{CID}^i$ can be written as follows:
\begin{equation}
\small
p(\mathrm{CID}^i|e_q,r_q,\theta )=\prod_{j=1}^{|\mathrm{CID}^i|} p(\mathrm{CID}^i_j|e_q,r_q,\mathrm{CID}^i_{<j},\theta) 
\label{equation: autoregressive}
\end{equation}
where $\theta$ is the parameters of Cluster Decoder $D_C$. We denote this probability as inter-cluster mapping score $S_{inter}(q,\mathrm{CID}^i)$, which characterizes the matching between $q$ and $\mathcal{D}_l$ under coarse-grained features. For a training pair ($q,d^+$), we use CrossEntropy loss to train GDR to achieve inter-cluster matching correctly:
\begin{equation}
\small
\mathcal{L}_{Inter}=-\mathrm{log}\ p(\mathrm{CID}(d^+)|E_Q(q),\theta_{D_C}) \ .
\label{equation: L_inter}
\end{equation}
Following NCI, we use the encoder of T5-base \citep{DBLP:conf/nips/BrownMRSKDNSSAA20} to initialize $E_Q$ and randomly initialized PAWA decoder (see \citet{DBLP:conf/nips/WangHWMWCXCZL0022} for details) as $D_C$.

\paragraph{Intra-cluster Matching}
To further achieve the intra-cluster mapping $f_{intra}: \mathrm{CID}^{1:k} \rightarrow d^{1:k}$, GDR applies the matching mechnism of calculating representation similarity for retrieval. 
Specifically, GDR leverages the Document Encoder $E_D$ trained in section \ref{subsec:CID} to extract the fine-grained features of candidate documents $d^{1:|\mathcal{D}_l|}$ into semantic representations $r_d^{1:|\mathcal{D}_l|} \in \mathbb{R}^d$ in prior. 
Then we pick out the $d^i$ belonging to the recalled clusters $\mathrm{CID}^{1:k}$ in the previous stage and calculate the intra-cluster mapping score between them and $q$ as follows:
\begin{equation}
\small
S_{intra}(q,d^i)=\mathrm{Sigmoid}(\mathrm{sim}(r_q,r_d^i)).
\label{equation: S_intra}
\end{equation}
where sim(·) denotes the inner product function. The Sigmoid function is used to map $S_{intra}$ into [0,1] to align with $S_{inter}$. 
NLL loss is used to train GDR for intra-cluster mapping ability:
\begin{equation}
\small
\mathcal{L}_{Intra}=-\mathrm{log}\frac{e^{sim(q,d^+)}}{e^{sim(q,d^+)}+\sum_{i}^{n}e^{sim(q,d^-_i)} }
\label{equation: L_intra}
\end{equation}
where $d^+$ and $d^-$ refer to documents relevant and irrelevant to $q$ respectively.
On this basis, the overall mapping score of $d^i$ is defined as: 
\begin{equation}
\small
S_{overall}(q,d^i)=S_{inter}(q,\mathrm{CID}(d^i)) + \beta * S_{intra}(q,d^i)
\label{equation: S_overall}
\end{equation}
where $\beta$ is a hyperparameter which we set as 1 by default. In the end, we take the Top-$k$ documents according to $S_{overall}$ as the final retrieval set $\mathcal{D}_c$.

\subsection{Construction of Memory-friendly CIDs}
\label{subsec:CID}
Considering the limited memory volume of the model, we are supposed to construct memory-friendly CIDs to ease the mapping $f_{intra}$.
% cluster identifiers (CIDs) with favorable hierarchical prior knowledge to alleviate the burden of memorizing the coarse-grained semantic structures of $D_l$. 
% The first thing we need to consider is what properties should the document cluster identifiers (CIDs) own to help GDR learn the mapping of queries to relevant clusters with limited memory volume. 

Ideally, we would like the CIDs corresponding to documents relevant to the same query to have similar prefixes. 
Such property can provide a mapping relationship between the query and CIDs with lower entropy, so as to alleviate the memorizing burden. 
% (property 1)
What's more, the total number of document clusters should be determined by the memory volume (model size) rather than the size of $\mathcal{D}_l$ to avoid exceeding the memorizing volume. 
% (property 2)
Based on these considerations, our strategy for generating CIDs is shown in Algorithm \ref{alg:CID}. 

\begin{algorithm}
\small
    \caption{Generating document cluster identifiers (CIDs).}\label{alg:CID}
        \begin{algorithmic}[1]
\Require Corpus $d^{1:|\mathcal{D}_l|}$, Document Encoder $E_D$, \newline
        Inter-cluster number $k$, Intra-cluster number $c$
\Ensure Document cluster identifiers $\mathrm{CID}^{1:|\mathcal{D}_l|}$
\State Encode $d^{1:|\mathcal{D}_l|}$ with $E_D$ to obtain document representations $X^{1:|\mathcal{D}_l|}$
\Function{G{\fontsize{9}{12}\selectfont ENERATE}CID{\fontsize{9}{12}\selectfont S}}{$X^{1:N}$}

\State $C^{1:k} \leftarrow Kmeans(X^{1:N})$
\State $L\leftarrow \varnothing$
\For{$i\gets 1, k$}
    \State $L_{current} \leftarrow [i]*|C^{i}|$
    \If{$|C^{i}|\ge c$}
        \State $L_{rest} \leftarrow$ G{\fontsize{9}{12}\selectfont ENERATE}CID{\fontsize{9}{12}\selectfont S}($C^{i}$)
    \Else
        \State $L_{rest} \leftarrow [0]*|C^{i}|$
    \EndIf
    \State $L_{cluster} \leftarrow $Concat$(L_{current},L_{rest})$
    \State $L \leftarrow L$.Append$(L_{cluster})$
\EndFor
\State ReorderToOriginal$(L,X^{1:N},C^{1:k})$
\State \textbf{Return} $L$
\EndFunction
\State $\mathrm{CID}^{1:|\mathcal{D}_l|} \leftarrow $ G{\fontsize{9}{12}\selectfont ENERATE}CID{\fontsize{9}{12}\selectfont S}($X^{1:|\mathcal{D}_l|}$)
        \end{algorithmic}
\end{algorithm}

To meet the first property, we finetuned ERNIE-2.0-base \citep{DBLP:conf/aaai/SunWLFTWW20} model following \citet{DBLP:conf/iclr/ZhangGS0DC22} on the training set \footnote{All experiments in this work were conducted on the Natural Questions dataset \citep{DBLP:journals/tacl/KwiatkowskiPRCP19}} and then used the finetuned document encoder as $E_D$ in Algorithm \ref{alg:CID}. Compared to previous studies \citep{DBLP:conf/nips/Tay00NBM000GSCM22,DBLP:conf/nips/WangHWMWCXCZL0022} using BERT \citep{DBLP:conf/naacl/DevlinCLT19} as $E_D$, our strategy can fully leverage the knowledge in the training set. To analyse the qualities of CIDs generated with different $E_D$, we calculated the average prefix overlap $O_{pre}$ of CIDs between the relevant documents for each query in the validation set $S_{val}$ as follows:
\begin{equation}
\setlength{\jot}{8pt}
\small
\begin{gathered}
% \begin{aligned}
O_{pre} = \frac{1}{|S_{val}|}\sum_{q\in{S_{val}}}\frac{1}{|\mathcal{D}_q|^2}\sum_{i=1}^{|\mathcal{D}_q|}\sum_{j=1}^{|\mathcal{D}_q|} o_{pre}(\mathrm{CID}_q^i,\mathrm{CID}_q^j)\\
o_{pre} (s_1,s_2) = |LCP(s_1,s_2)|/|s_1|
% \end{aligned}
\end{gathered}
\label{equation: prefix_overlap}
\end{equation}
where $\mathrm{CID}_q^i$ is the cluster identifier of the $i^{th}$ relevant document of $q$ and $LCP(s_1,s_2)$ is the longest common prefix of string $s_1$ and $s_2$. The results show that the $O_{pre}$ corresponding to the CIDs generated by our strategy (0.636) is significantly higher than the previous study (0.516), indicating that our CIDs is more distinguishable and can better meet the first property . To meet the second property, we consider adaptively changing $c$ in Algorithm \ref{alg:CID} to ensure the total number of clusters $|\mathrm{CID}|$ not to change with $\mathcal{D}_l$ as follows:
\begin{equation}
\small
c = |\mathcal{D}_l| \ / \  Exp(|\mathrm{CID}|)
\end{equation}
where $Exp(|\mathrm{CID}|)$ is the expected value of $|\mathrm{CID}|$ which we set as 5000 in our experiment for simplicity. Under different sizes of $\mathcal{D}_l$, the $|\mathrm{CID}|$ we obtained through this strategy is basically in the same order of magnitude (Appendix \ref{sec: Magnitude of CIDs}), which meets the second properties.

\subsection{Cluster-adaptive Negative Sampling}
An important issue in calculating $\mathcal{L}_{Intra}$ is how to select $d^-$ with effective training signals. Various negative sampling methods (e.g., static bm25-based sampling \citep{DBLP:conf/emnlp/KarpukhinOMLWEC20}, dynamic index-based sampling \citep{DBLP:conf/iclr/XiongXLTLBAO21}) have been proposed to pick up hard negatives. However, GDR needs to retrieve relevant documents within the candidate clusters instead of the entire corpus, which requires negative samples to offer more intra-cluster discriminative signals. To this end, we propose cluster-adaptive negative sampling strategy. For a training pair ($q,d^+$), we treat $d \in{\mathrm{CID}(d^+)}$ as intra-cluster negatives $N_a$ and in-batch negatives \citep{DBLP:journals/corr/HendersonASSLGK17} as inter-cluster negatives $N_r$, and rewrite Eq.~\eqref{equation: L_intra} as follows:

\begin{equation}
\small
\mathcal{L}_{Intra}=-\mathrm{log}\frac{e^{\mathrm{sim}(q,d^+)}}{\gamma*\sum_{d\in{N_a}}e^{\mathrm{sim}(q,d)}+\sum_{d\in{N_r}}e^{\mathrm{sim}(q,d)} }
\label{equation: L_intra_final}
\end{equation}
where $\gamma$ is a hyperparameter we set as 2 to enhance intra-cluster discriminative training signals.

\subsection{Training and Inference}
\paragraph{Training Phase}
Given a corpus $\mathcal{D}_l$ and a training set $\mathcal{S}_{train} = \{(q^i,d^i)|i\in{(1,...,n)}\}$, we use DocT5Query \footnote{\url{https://github.com/castorini/docTTTTTquery}} to generate 5 pseudo queries through and randomly select 5 groups of 40 consecutive terms from the document as additional queries for each document. Compared with \citet{DBLP:conf/nips/WangHWMWCXCZL0022} that augment each document with totally 26 queries, fewer augmented queries are required as GDR only needs to memorize coarse-grained semantics, thus saves training expenses. The augmented training set $\mathcal{S}_{aug}$ together with $\mathcal{S}_{train}$ are used to train GDR using the total loss:
\begin{equation}
\small
\mathcal{L}_{GDR}=\mathcal{L}_{Inter}+\mathcal{L}_{Intra}
\label{equation: L_total}
\end{equation}
To accelerate the training process, we use $E_D$ to calculate the representations of $\mathcal{D}_l$ in advance and freeze the parameters of $E_D$ during training phase.

\paragraph{Inference Phase}
During inference, we first generate k relevant CIDs through beam search, and then retrieve the top-m documents with highest $S_{intra}$ in each relevant cluster (m is the minimum value between the number of documents in the cluster and k). Finally, we reorder all these documents according to $S_{overall}$ to obtain the most relevant top-k documents. Following \citet{DBLP:conf/nips/Tay00NBM000GSCM22}, we pre-build a prefix tree to ensure only the valid CIDs can be generated.  
We conduct Approximate Nearest Neighbor Search \citep{DBLP:journals/tkde/LiZSWLZL20} in each cluster to accelerate the intra-cluster matching process.

\section{Experiments}
We empirically demonstrate the performance of GDR and effectiveness of various proposed strategies on text retrieval task in this section.\footnote{we will release our code as soon as the paper is accepted} In the following, we will discuss the detailed experimental setups in \ref{subsec:Experimental settings}, present empirical results in \ref{subsec:Main Results}, verify the effectiveness of proposed modules in \ref{subsec:Ablation}, and conduct specific analysis in \ref{subsec:Analysis}, respectively. 
%where the documents are gathered from Wikipedia pages and the queries are natural language questions

\subsection{Experimental settings}
\label{subsec:Experimental settings}
    \paragraph{Datasets}
    % We conduct all the experiments on the challenging Natural Questions (NQ) \citep{DBLP:journals/tacl/KwiatkowskiPRCP19} dataset downloaded from Tevatron\footnote{https://huggingface.co/Tevatron}. 
    We choose classic text retrieval dataset Natural Questions \footnote{We use the cleaned version of NQ downloaded from \url{https://huggingface.co/Tevatron}} (NQ) \citep{DBLP:journals/tacl/KwiatkowskiPRCP19} for experiment, which consists of 58K (query, relevant passages) training pairs and 6K validation pairs along with 21M candidate passage corpus. 
    % The length of passages is limited to 100 tokens and the queries are natural language questions. 
    Each query corresponds to an average of 7.5 relevant passages, which puts higher demands on the recall performance of the model. We gather all the relevant passages of queries included in NQ training and validation set, resulting in a 334K candidate passage corpus setting (NQ334K). 
    We further build NQ1M, NQ2M, and NQ4M settings to evaluate the performance of GDR on larger corpus by adding the remaining passages from the full 21M corpus to NQ334K.
    % To evaluate the performance of GDR on larger corpus setting, we continuously add the remaining documents from the NQ document corpus to build NQ1M, NQ2M, and NQ4M settings, which indicates different numbers of total candidate documents in the dataset. 
    For GDR, CIDs are generated separately for each dataset so as to prevent leakage of semantic information from larger candidate document corpus into smaller ones. GDR of different settings are trained on the training set together with corresponding augmented set, and evaluated on the validation set \footnote{
    The lack of relevant documents makes the test set inconvenient to partition different settings}.

    \begin{table*}
    \renewcommand\arraystretch{1.2}
    \small
    \centering
    \setlength{\tabcolsep}{0.195em} % 设置列之间的间距
    \begin{tabular}{cl|cc|cc|cc|cc}
    \hline
    \multirow{2}{*}{Paradigm} & \multirow{2}{*}{Method} & \multicolumn{2}{c|}{NQ334K} & \multicolumn{2}{c|}{NQ1M} & \multicolumn{2}{c|}{NQ2M} & \multicolumn{2}{c}{NQ4M} \\
     & & Acc@20/100 & R@20/100 & Acc@20/100 & R@20/100 & Acc@20/100 & R@20/100 & Acc@20/100 & R@20/100\\   

    \cline{1-10}
    SR & BM25 & 86.1 / 92.4 & 56.0 / 75.4 & 84.0 / 91.0 & 51.3 / 73.0 & 82.4 / 89.9 & 47.5 / 71.0 & 79.6 / 88.4 & 42.3 / 68.2 \\
    % & SPLADE & 91.4 / 95.8 & 58.8 / 75.6 & 89.9 / 94.2 & 55.2 / 73.3 & 88.8 / 94.2 & 52.9 / 71.2 & 87.0 / 93.1 & 49.5 / 68.5\\
    \cline{1-10}
    \multirow{2}{*}{DR} & DPR & 93.9 / 97.3 & 49.8 / 60.2 & 91.5 / 96.3 & 46.7 / 56.6 & 90.4 / 95.5 & 45.2 / 54.9 & 88.4 / 94.6 & 42.9 / 52.8 \\
    & AR2 & \textbf{96.3} / \textbf{98.6} & 57.4 / 69.0 & \textbf{94.9} / \textbf{98.0} & 54.7 / 66.2 & \textbf{94.3} / \textbf{97.7} & 53.2 / 64.7 & \textbf{93.4} / \textbf{97.2} & 51.2 / 62.6 \\
    \cline{1-10}
    \multirow{2}{*}{GR} & NCI-bert & 80.0 / 88.7 & 49.4 / 65.5 & 72.0 / 82.6 & 38.7 / 54.5 & 63.9 / 76.4 & 30.2 / 44.6 & 55.4 / 70.0 & 25.2 / 37.8\\
    & NCI-ours & 88.0 / 94.1 & 60.0 / 75.6 & 80.3 / 89.6 & 50.6 / 66.2 & 78.2 / 88.6 & 46.4 / 63.5 & 77.3 / 87.8 & 45.2 / 61.0\\
    \cline{1-10}
    \multirow{2}{*}{GDR} & GDR-bert & 87.5 / 91.2 & 59.3 / 71.2 & 84.8 / 88.8 & 54.8 / 66.0 & 83.3 / 88.0 & 51.9 / 64.8 & 82.1 / 87.7 & 49.7 / 63.8 \\
    & GDR-ours & 91.1 / 95.3 & \textbf{64.6} / \textbf{79.6} & 88.2 / 93.6 & \textbf{60.1} / \textbf{75.2} & 87.4 / 92.8 & \textbf{57.7} / \textbf{73.2} & 87.0 / 92.2 & \textbf{55.2} / \textbf{71.5}\\
    \hline
    \end{tabular}
    \caption{Experimental results on NQ document retrieval. The settings "-bert" and "-ours" denote using BERT and our finetuned $E_D$ in section \ref{subsec:CID} to generate document embeddings for the generation of identifiers respectively. Bold numbers represent best performance. We run four random seeds and report the averaged result for each method.}
    \label{tab: Main results}
    \end{table*}

    \paragraph{Evaluation metrics}
    We use widely accepted metrics for text retrieval, including $\mathrm{R}@k$ (also denoted as $\mathrm{Recall}@k$) and $\mathrm{Acc}@k$, where $k\in{\{20, 100\}}$. Specifically, $\mathrm{R}@k$ calculates the proportion of relevant documents included in top-$k$ retrieved candidates ($\# \operatorname{retr}_{q,k}$) among all the candidate relevant documents ($\# \operatorname{rel}_q$) (Eq.~\eqref{equation: recall}), while $\mathrm{Acc}@k$ measures how often the correct document is hit by top-$k$ retrieved candidates (Eq.~\eqref{equation: acc}).
    \begin{equation}
    \small
    \mathrm{R}@k=\frac{1}{|\mathcal{S}_{val}|}\sum_{q\in{\mathcal{S}_{val}}}\frac{\# \operatorname{retr}_{q,k}}{\# \operatorname{rel}_q}
    \label{equation: recall}
    \end{equation}
    \begin{equation}
    \small
    \mathrm{Acc}@k=\frac{1}{|\mathcal{S}_{val}|}\sum_{q\in{\mathcal{S}_{val}}}\mathbb{I}\left(\# \operatorname{retr}_{q, k}>0\right)
    \label{equation: acc}
    \end{equation}
    \paragraph{Baselines}
    We choose the following methods for detailed comparisons. BM25 (Anserini implementation \citep{DBLP:conf/sigir/Yang0L17}) 
    % and SPLADE v2 \footnote{\url{https://github.com/naver/splade}} \citep{DBLP:journals/corr/abs-2109-10086} are 
    is served as a strong \textbf{SR} baseline. As for \textbf{DR}, we select a strong baseline DPR \footnote{\url{https://github.com/facebookresearch/DPR}} \citep{DBLP:conf/emnlp/KarpukhinOMLWEC20} and state-of-the-art (SOTA) method AR2 \footnote{\url{https://github.com/microsoft/AR2}} \citep{DBLP:conf/iclr/ZhangGS0DC22}. As for \textbf{GR}, we select the SOTA method NCI \footnote{\url{https://github.com/solidsea98/Neural-Corpus-Indexer-NCI}} \citep{DBLP:conf/nips/WangHWMWCXCZL0022}. To ensure the reliability of the experimental results, we reproduce all the baseline methods based on their official implementations.
    % \footnote{\url{}}  https://github.com/facebookresearch/DPR \\ https://github.com/microsoft/AR2 \\ https://github.com/solidsea98/Neural-Corpus-Indexer-NCI}
    % We evaluate BM25(Anserini implementation\citep{DBLP:conf/sigir/Yang0L17}) and DPR \citep{DBLP:conf/emnlp/KarpukhinOMLWEC20} on our testbed as they are strong baselines in SR and DR methods respectively. To further demonstrate the effectiveness of GDR, we consider state-of-the-art(SOTA) method AR2\citep{DBLP:conf/iclr/ZhangGS0DC22} in DR and SOTA method NCI\citep{DBLP:conf/nips/WangHWMWCXCZL0022} in GR for comparison. To ensure the reliability of the experimental results, we reproduce all the baseline methods by their official implementations.\footnote{\footnotesize https://github.com/solidsea98/Neural-Corpus-Indexer-NCI \\ https://github.com/facebookresearch/DPR \\ https://github.com/microsoft/AR2}
    \paragraph{Experimental details}
    We implement GDR with python 3.8.12, PyTorch 1.10.0 and HuggingFace transformers 3.4.0. The learning rates are set as  $2\times 10^{-4}$ for the Query Encoder and $1\times 10^{-4}$ for the Cluster Decoder with a batch size 256 per GPU. For inference, we apply the constraint beam search algorithm, and set the length penalty and the beam size as 0.8 and 100, respectively. All experiments are based on a cluster of NVIDIA A100 GPUs with 40GB memory. Each job takes 8 GPUs, resulting in a total batch size of 2048 (256 $\times$ 8). We train the GDR models for 60 epochs and pick the final checkpoint for evaluation.

\subsection{Main Results}
\label{subsec:Main Results}
    \paragraph{Horizontal Comparison}
    As shown in the Table \ref{tab: Main results}, the performance of each method on $\mathrm{R}@k$ metrics is as follows: GDR (GDR-ours) > SR (BM25) > DR (AR2) > GR (NCI), while the ranking on $\mathrm{Acc}@k$ metrics is as follows: DR (AR2)  > GDR (GDR-ours) > SR (BM25) > GR (NCI). Based on the characteristics of sparse lexical matching, SR can recall the majority of relevant documents ($2nd$ $\mathrm{R}@k$) when the query is accurate while may not even hit one target when there is a lexical mismatch ($3rd$ $\mathrm{Acc}@k$). On the contrary, DR can hit at least one relevant document in most situations by semantic representation matching ($1st$ $\mathrm{Acc}@k$). However, the semantic differences in relevant documents make it difficult to recall them all simultaneously ($3rd$ $\mathrm{R}@k$). GR (NCI) ranks last due to the difficulty in memorizing large-scale corpus we have discussed. 

    By conducting a coarse-to-fine retrieval process, GDR maximizes the advantages of memorizing mechanism in deep interaction and matching mechanism in fine-grained features discrimination, thus ranks $1st$ on $\mathrm{R}@k$ with an average of 3.0 improvement and $2nd$ on $\mathrm{Acc}@k$.

    % due to the feature of semantic representation matching, DR can hit most relevant documents(highest Accuracy) while performing poor on recalling all the relevant documents, whose features can be sparse in the semantic space(low Recall); as a result of the larger corpus memory burden, the performance of GR (NCI) significantly drop on both Recall metrics and Accuracy metrics compared to the SOTA result reported in NCI\citep{DBLP:conf/nips/WangHWMWCXCZL0022}; As a comparison, GDR exhibits good performance on both Recall and Accuracy metrics, specifically by: outperforming the best baseline model on recall metrics by $+8.6\%$ $\mathrm{R}@20$ and $+4.2\%$ $\mathrm{R}@100$ and achieves relatively high performance(ranking second among all the methods) on Accuracy metrics, which verifies the superiority of GDR over SR, DR and GR methods. 
    % Compared to DR,  . On the small candidate corpus dataset(NQ334K), 
    % For Acc@20 and Acc@100 metrics, we notice that DR method outperforms SR, GR and GDR methods across all datasets, and the proposed GDR ranks second among all the methods.

\begin{figure}[t]
    \centering
    \includegraphics[width=1.0\linewidth]{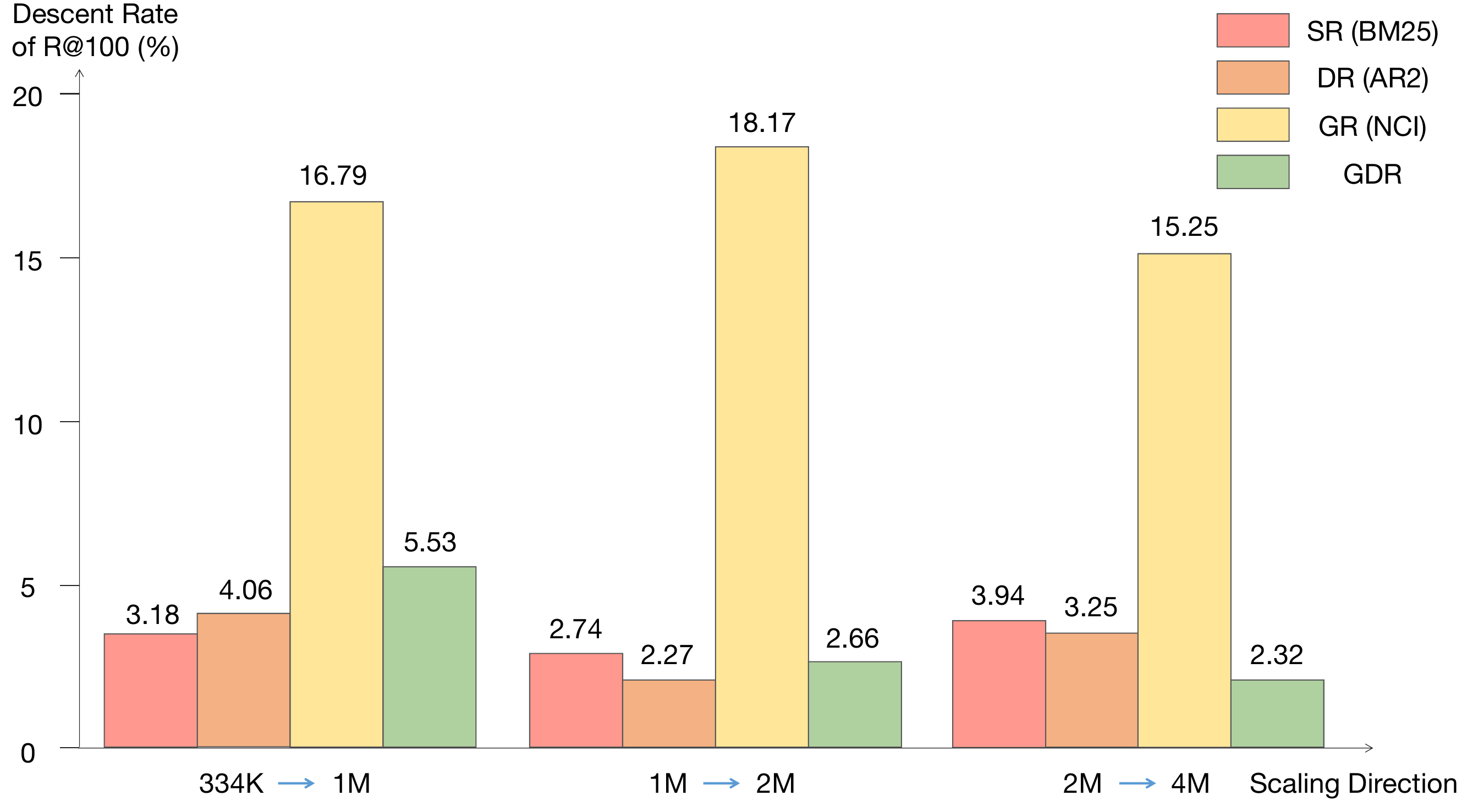}
    \caption{$\mathrm{R}@100$ descent rate of different types of methods when scaling to larger corpus.}
    \label{fig: scaling}
    % \vspace{-1.0\baselineskip}nnnj
\end{figure}
    
    \paragraph{Scaling to Larger Corpus} Memorizing mechanism has been proven to bring advanced retrieval performance under small corpus settings \citep{DBLP:conf/nips/WangHWMWCXCZL0022}. However, when the corpus size that needs to be memorized exceeds the memory volume, it can instead become a burden. As shown in Figure \ref{fig: scaling}, when the candidate corpus scaling to larger size, the descent rate of $\mathrm{R}@100$ for both SR and DR keeps below $4.06\%$, while it astonishingly retains exceeding $15.25\%$ for GR on all three scaling directions. As a comparison, GDR ensures the maximum utilization of memorizing mechanism by focusing memory content on fixed volume coarse-grained features of corpus to achieve inter-cluster matching. This strategy results in GDR achieving an average of $3.50\%$ descent rate of $\mathrm{R}@100$, which is almost the same as SR ($3.29\%$) and DR ($3.19\%$).

\subsection{Ablation studies on Model training}
\label{subsec:Ablation}
    To further understand how different paradigm options affect model performance, we conduct ablation experiments and discuss our findings below.
\paragraph{Cluster Identifiers} 
    We first analyse the influence of identifiers constructed with documents representations generated by different models. Specifically, the results are shown in Table \ref{tab: Main results}, where "-bert" and "-ours" denotes using BERT and our finetuned model as $E_D$ in Algorithm \ref{alg:CID} respectively. Basically, both NCI and GDR trained with "-ours" perform significantly better than those trained with "-bert" across all the settings. The results empirically demonstrate that fully leveraging the knowledge in the training set to generate identifiers that characterizing a mapping from query to relevant documents with lower entropy can significantly release the memorizing burden thus leading to better retrieval performances. Considering that NCI has a heavier memory burden compared to GDR, this strategy has benefited NCI more (10.1 > 8.4 $\mathrm{R}@100$ improvements on NQ334K).
    
    % In the setting of fixed total number of document clusters, we evaluate the performance of model trained with identifiers generated by BERT(denote as "-bert") and  finetuned ERNIE-2.0-base(denote as "-ar2"), respectively. The results are shown in Table \ref{tab: Main results}. We see that, the model trained with "-ar2" performs significantly better than those trained with "-bert" across all the datasets. On NQ334K dataset, NCI-ar2 outperforms NCI by $+5.4\%$ $\mathrm{Acc}@100$ and $+10.1\%$ $\mathrm{R}@100$, while GDR-ar2 outperforms GDR-bert by $+5.3\%$ $\mathrm{Acc}@100$ and $+8.4\%$ $\mathrm{R}@100$. Furthermore, the performance gap between NCI and NCI-ar2 becomes larger as the corpus size grows. The results demonstrate that, compared to the pretrained model BERT, using our finetuned $E_D$ to generate features of documents can fully leverage the knowledge in the training set, and provide better knowledge for the generation of indentifiers, thus enables the model to learn a more superior mapping route from query to the relevant documents.

\paragraph{Negative Sampling Strategy}
    To verify the effectiveness of the proposed cluster-adaptive negative sample strategy, We evaluate the performance of GDR trained with different negative sampling strategies and summarize the results in Table \ref{tab: Negative sample strategies}. We notice that GDR trained with the cluster-adaptive strategy outperforms that with widely used BM25 strategy by 1.1 on $\mathrm{R}@100$. This indicates that our proposed cluster-adaptive negative sampling strategy can indeed provide more intra-cluster discriminative training signals to strengthen the fine-grained matching ability.
    
\begin{table}[t]
    \small
    \centering
    \setlength{\tabcolsep}{0.45em} % 设置列之间的间距
    \begin{tabular}{lcccc}
    \toprule
    %\hline
    Strategy & Acc@20 & Acc@100 & R@20 & R@100\\   
    \midrule
    Random & 87.1 & 91.4 & 60.8 & 76.0\\
    BM25 & 90.2 & 94.6 & 63.1 & 78.5\\
    Cluster-adaptive & \textbf{91.1} & \textbf{95.3} & \textbf{64.6} & \textbf{79.6}\\
    %\cline{1-10}
    \bottomrule
    \end{tabular}
    \caption{Comparison of the performance of GDR trained with different negative sample strategies on NQ334K dataset. }
    \label{tab: Negative sample strategies}
\end{table}

\begin{table}[t]
    \renewcommand\arraystretch{1.2}
    \small
    \centering
    \begin{tabular}{lcccc}
    \toprule[1pt]
    $\beta$ & Acc@20 &Acc@100 & R@20 & R@100\\
    \hline
    0 & 70.5 & 83.9 &39.2 & 59.4 \\
    0.5 & 89.1 &93.7 & 61.9 & 77.2 \\
    1 & \textbf{91.1} & \textbf{95.3} &\textbf{64.6} & \textbf{79.6} \\
    2 & 90.9 & 95.0 & 64.4 & 79.5 \\
    1e5 & 90.4 & 94.8 & 63.1 & 77.9 \\
    \bottomrule[1pt]
    \end{tabular}
    \caption{Results of GDR with different $\beta$ on NQ334K dataset.}
    \label{tab: beta}
\end{table}

\subsection{Analysis}
\label{subsec:Analysis}
\paragraph{Combination of Mapping Scores}
We study the influence of different combination weights of $S_{inter}$ and $S_{intra}$ in Eq.~\eqref{equation: S_overall} and choose the value of $\beta$ from \{0,0.5,1,2,1e5\}. As the beta gradually increases (Table \ref{tab: beta}), the retrieval performance of GDR will experience a process of first increasing and then decreasing. Therefore, we take the best performing ($\beta$=1) as the default setting. When GDR only relies on $S_{inter}$ for retrieval ($\beta$ = 0), the ranking of documents within the same cluster will be the same, which will result in a significant performance degradation compared with the default setting. On the contrary, when GDR only relies on $S_{intra}$ for retrieval (we set $\beta$ = 1e5 to approximate this situation), the lack of matching information of coarse-grained semantic features will result in a decrease of 1.7 $\mathrm{R}@100$. The above experimental results fully demonstrate the significance of $S_{inter}$ and $S_{intra}$ and the necessity of combining them.

\begin{table}[t]
    \small
    \centering
    \setlength{\tabcolsep}{0.19em} % 设置列之间的间距
    \begin{tabular}{lllllll}
    \hline
    \multirow{2}{*}{$\mathcal{D}_l$}& \multirow{2}{*}{$\mathcal{S}_{val}$} & \multicolumn{2}{c}{NCI} & & \multicolumn{2}{c}{GDR} \\
    \cline{3-4} \cline{6-7} & & Acc@100 & R@100 & & Acc@100 & R@100 \\   
    \hline
    Set A & Set A   & 90.7\hspace{0.30cm}- & 71.2\hspace{0.30cm}- & & 94.9\hspace{0.30cm}- & 77.7\hspace{0.30cm}- \\
    All & Set A  & 80.7\textcolor{lightgreen}{$\downarrow$10.0} & 52.9\textcolor{lightgreen}{$\downarrow$18.3} & & 93.4\textcolor{lightgreen}{$\downarrow$1.0\hspace{0.16cm}} & 75.8\textcolor{lightgreen}{$\downarrow$1.9\hspace{0.16cm}} \\
    All & Set B & 56.5\textcolor{lightgreen}{$\downarrow$34.2} & 27.7\textcolor{lightgreen}{$\downarrow$43.5} & & 86.6\textcolor{lightgreen}{$\downarrow$8.3} & 66.2\textcolor{lightgreen}{$\downarrow$11.5}\\ 
    \hline
    \end{tabular}
    \caption{Comparison of scalability performance between NCI and GDR. Specifically, We divide the original NQ334K dataset into two parts: Set A (constructing identifiers and training on it) and Set B (served as new added dataset). }
    \label{tab: scalability of model}
\end{table}

\paragraph{Scalability of Model}
A common scenario in retrieval tasks is adding new documents to candidate corpus. To simulate this scenario, we split the NQ334K dataset into Set A and Set B, both of which contain half of the original training and validation set together with corresponding relevant documents. For both NCI and GDR, we first train and evaluate the model on Set A. After adding Set B to Set A, we further evaluate the model on validation subset of Set A and Set B respectively. As shown in Table \ref{tab: scalability of model}, though NCI has already memorized the documents corresponding to Set A validation set, the situations where one document identifier corresponds to multiple documents caused by the new added documents led to a 18.3 $\mathrm{R}@100$ drop. On the contrary, GDR only degraded 1.9 on $\mathrm{R}@100$ thanks to the introduction of $S_{intra}$. When evaluating on Set B, NCI further significantly degraded 25.2 on $\mathrm{R}@100$ as the model did not have a memory of documents corresponding to Set B validation set. 
As a comparison, GDR can quickly extract dense representations through $E_D$ and assign cluster identifiers by searching for the nearest cluster representation in the semantic space for the added documents, so as to obtain inter-cluster and intra-cluster features. Although GDR also does not have a memory of added documents, its $\mathrm{R}@100$ performance (66.2) still significantly surpassed NCI (27.7) on Set B.

\begin{table}[t]
    \renewcommand\arraystretch{1.2}
    \small
    \centering
    \begin{tabular}{lccc}
    \toprule[1pt]
    \multirow{2}{*}{Method} & Latency & Throughput & Index Refresh \\
    &(ms)&(queries/s)&(mins)\\
    \hline
    BM25  & 56 & 22.8&2 \\
    AR2  & 35 & 589.0&5 \\ 
    NCI & 232 & 6.3&- \\
    GDR  & 195 & 7.2&7 \\
    \bottomrule[1pt]
    \end{tabular}
    \caption{Efficiency analysis on NQ334K dataset with recall quantity as 100. NCI can not refresh indexes without retraining.}
    \label{tab: efficiency}
\end{table}

\paragraph{Efficiency Analysis}
We use an NVIDIA A100-40G GPU to analyze the efficiency of AR2, NCI, and GDR. We use the Anserini implementation of BM25 and evaluate it on an Intel Xeon CPU. As shown in Table \ref{tab: efficiency}, BM25 and AR2 achieve fast retrieval by indexing the corpus in advance. Typical GR method NCI has lower efficiency due to the autoregressive generation of document identifiers with beam search. As a compromise, GDR uses autoregressive generation in inter-cluster matching and pre-indexes for retrieval in intra-cluster matching, thus achieves an efficiency that falls between DR and GR. We leave the research on improving the efficiency of GR and GDR for future work.
%Ablation analysis here

% Considering the limited memory volume of the model, we are supposed to construct cluster identifiers (CIDs) with favorable hierarchical prior knowledge to alleviate the burden of memorizing the coarse-grained semantic structures of $D_l$. 

\section{Conclusions}
In this paper, we empirically demonstrate that the memorizing mechanism of Generative Retrieval (GR) brings deep interaction characteristics but also causes serious problems. 
To this end, we propose the Generative Dense Retrieval (GDR) paradigm, which subdivides the text retrieval task into inter-cluster and intra-cluster matching and achieves them by autoregressively generating cluster identifiers and calculating dense representation similarities respectively. 
% By concentrating on memorizing the coarse-grained semantics of corpus, GDR can maximize the deep interaction and multi-directions matching between query and documents, thus achieving good recall performance. 
% The introduction of dense representation captures precise fine-grained features of documents to achieve accurate intra-cluster matching and also brings better scalability. 
GDR focuses the limited memory volume on the deep interaction between query and document cluster and conducts multi-directions decoding, thus maintaining the superiority of memorizing mechanism.
Memorizing-free matching mechanism is further introduced to achieve intra-cluster mapping by fully leveraging fine-grained features of documents.
Such a coarse-to-fine process can also bring better scalability, i.e., stable corpus expansion and low-cost document updates. 
We further propose a cluster identifier constructing strategy to release the memory burden and a cluster-adaptive negative sampling strategy to provide discriminative signals.
Comprehensive experiments on the NQ dataset demonstrate the state-of-the-art $\mathrm{R}@k$ performance and better scalability of GDR.

\section*{Limitations}
Despite the achievement of state-of-the-art $\mathrm{R}@k$ performance and better scalability, the current implementation of GDR still suffers from the following limitations. Firstly, the inference speed of GDR needs to be further improved to be employed in real-time retrieval services. Secondly, GDR's performance on $\mathrm{Acc}@k$ falls short compared to the state-of-the-art method (AR2 \citep{DBLP:conf/iclr/ZhangGS0DC22}). We suppose that this is because part of the Query Encoder's capacity is utilized to handle the inter-cluster matching task, thus affects the accuracy of GDR in intra-cluster mapping. Thirdly, due to the high training cost (70 hours on 8 NVIDIA A100 GPUs for NQ4M), the generalization of GDR on larger scale corpus has not been tested.

\section*{Ethics Statement}
All of the datasets used in this study were publicly available, and no annotators were employed for data collection. We confirm that the datasets we used did not contain any harmful content and was consistent with their intended use (research). We have cited the datasets and relevant works used in this study. 

\bibliography{acl_latex}

\begin{thebibliography}{26}
\expandafter\ifx\csname natexlab\endcsname\relax\def\natexlab#1{#1}\fi

\bibitem[{Bevilacqua et~al.(2022)Bevilacqua, Ottaviano, Lewis, Yih, Riedel, and Petroni}]{DBLP:conf/nips/BevilacquaOLY0P22}
Michele Bevilacqua, Giuseppe Ottaviano, Patrick S.~H. Lewis, Scott Yih, Sebastian Riedel, and Fabio Petroni. 2022.
\newblock \href {http://papers.nips.cc/paper\_files/paper/2022/hash/cd88d62a2063fdaf7ce6f9068fb15dcd-Abstract-Conference.html} {Autoregressive search engines: Generating substrings as document identifiers}.
\newblock In \emph{NeurIPS}.

\bibitem[{Brickley et~al.(2019)Brickley, Burgess, and Noy}]{DBLP:conf/www/BrickleyBN19}
Dan Brickley, Matthew Burgess, and Natasha~F. Noy. 2019.
\newblock \href {https://doi.org/10.1145/3308558.3313685} {Google dataset search: Building a search engine for datasets in an open web ecosystem}.
\newblock In \emph{The World Wide Web Conference, {WWW} 2019, San Francisco, CA, USA, May 13-17, 2019}, pages 1365--1375. {ACM}.

\bibitem[{Brown et~al.(2020)Brown, Mann, Ryder, Subbiah, Kaplan, Dhariwal, Neelakantan, Shyam, Sastry, Askell, Agarwal, Herbert{-}Voss, Krueger, Henighan, Child, Ramesh, Ziegler, Wu, Winter, Hesse, Chen, Sigler, Litwin, Gray, Chess, Clark, Berner, McCandlish, Radford, Sutskever, and Amodei}]{DBLP:conf/nips/BrownMRSKDNSSAA20}
Tom~B. Brown, Benjamin Mann, Nick Ryder, Melanie Subbiah, Jared Kaplan, Prafulla Dhariwal, Arvind Neelakantan, Pranav Shyam, Girish Sastry, Amanda Askell, Sandhini Agarwal, Ariel Herbert{-}Voss, Gretchen Krueger, Tom Henighan, Rewon Child, Aditya Ramesh, Daniel~M. Ziegler, Jeffrey Wu, Clemens Winter, Christopher Hesse, Mark Chen, Eric Sigler, Mateusz Litwin, Scott Gray, Benjamin Chess, Jack Clark, Christopher Berner, Sam McCandlish, Alec Radford, Ilya Sutskever, and Dario Amodei. 2020.
\newblock \href {https://proceedings.neurips.cc/paper/2020/hash/1457c0d6bfcb4967418bfb8ac142f64a-Abstract.html} {Language models are few-shot learners}.
\newblock In \emph{Advances in Neural Information Processing Systems 33: Annual Conference on Neural Information Processing Systems 2020, NeurIPS 2020, December 6-12, 2020, virtual}.

\bibitem[{Cao et~al.(2021)Cao, Izacard, Riedel, and Petroni}]{DBLP:conf/iclr/CaoI0P21}
Nicola~De Cao, Gautier Izacard, Sebastian Riedel, and Fabio Petroni. 2021.
\newblock \href {https://openreview.net/forum?id=5k8F6UU39V} {Autoregressive entity retrieval}.
\newblock In \emph{9th International Conference on Learning Representations, {ICLR} 2021, Virtual Event, Austria, May 3-7, 2021}. OpenReview.net.

\bibitem[{Chen et~al.(2017)Chen, Liu, Yin, and Tang}]{DBLP:journals/sigkdd/ChenLYT17}
Hongshen Chen, Xiaorui Liu, Dawei Yin, and Jiliang Tang. 2017.
\newblock \href {https://doi.org/10.1145/3166054.3166058} {A survey on dialogue systems: Recent advances and new frontiers}.
\newblock \emph{{SIGKDD} Explor.}, 19(2):25--35.

\bibitem[{Devlin et~al.(2019)Devlin, Chang, Lee, and Toutanova}]{DBLP:conf/naacl/DevlinCLT19}
Jacob Devlin, Ming{-}Wei Chang, Kenton Lee, and Kristina Toutanova. 2019.
\newblock \href {https://doi.org/10.18653/v1/n19-1423} {{BERT:} pre-training of deep bidirectional transformers for language understanding}.
\newblock In \emph{Proceedings of the 2019 Conference of the North American Chapter of the Association for Computational Linguistics: Human Language Technologies, {NAACL-HLT} 2019, Minneapolis, MN, USA, June 2-7, 2019, Volume 1 (Long and Short Papers)}, pages 4171--4186. Association for Computational Linguistics.

\bibitem[{Henderson et~al.(2017)Henderson, Al{-}Rfou, Strope, Sung, Luk{\'{a}}cs, Guo, Kumar, Miklos, and Kurzweil}]{DBLP:journals/corr/HendersonASSLGK17}
Matthew~L. Henderson, Rami Al{-}Rfou, Brian Strope, Yun{-}Hsuan Sung, L{\'{a}}szl{\'{o}} Luk{\'{a}}cs, Ruiqi Guo, Sanjiv Kumar, Balint Miklos, and Ray Kurzweil. 2017.
\newblock \href {http://arxiv.org/abs/1705.00652} {Efficient natural language response suggestion for smart reply}.
\newblock \emph{CoRR}, abs/1705.00652.

\bibitem[{Karpukhin et~al.(2020)Karpukhin, Oguz, Min, Lewis, Wu, Edunov, Chen, and Yih}]{DBLP:conf/emnlp/KarpukhinOMLWEC20}
Vladimir Karpukhin, Barlas Oguz, Sewon Min, Patrick S.~H. Lewis, Ledell Wu, Sergey Edunov, Danqi Chen, and Wen{-}tau Yih. 2020.
\newblock \href {https://doi.org/10.18653/v1/2020.emnlp-main.550} {Dense passage retrieval for open-domain question answering}.
\newblock In \emph{Proceedings of the 2020 Conference on Empirical Methods in Natural Language Processing, {EMNLP} 2020, Online, November 16-20, 2020}, pages 6769--6781. Association for Computational Linguistics.

\bibitem[{Kwiatkowski et~al.(2019)Kwiatkowski, Palomaki, Redfield, Collins, Parikh, Alberti, Epstein, Polosukhin, Devlin, Lee, Toutanova, Jones, Kelcey, Chang, Dai, Uszkoreit, Le, and Petrov}]{DBLP:journals/tacl/KwiatkowskiPRCP19}
Tom Kwiatkowski, Jennimaria Palomaki, Olivia Redfield, Michael Collins, Ankur~P. Parikh, Chris Alberti, Danielle Epstein, Illia Polosukhin, Jacob Devlin, Kenton Lee, Kristina Toutanova, Llion Jones, Matthew Kelcey, Ming{-}Wei Chang, Andrew~M. Dai, Jakob Uszkoreit, Quoc Le, and Slav Petrov. 2019.
\newblock \href {https://doi.org/10.1162/tacl\_a\_00276} {Natural questions: a benchmark for question answering research}.
\newblock \emph{Trans. Assoc. Comput. Linguistics}, 7:452--466.

\bibitem[{Li et~al.(2020)Li, Zhang, Sun, Wang, Li, Zhang, and Lin}]{DBLP:journals/tkde/LiZSWLZL20}
Wen Li, Ying Zhang, Yifang Sun, Wei Wang, Mingjie Li, Wenjie Zhang, and Xuemin Lin. 2020.
\newblock \href {https://doi.org/10.1109/TKDE.2019.2909204} {Approximate nearest neighbor search on high dimensional data - experiments, analyses, and improvement}.
\newblock \emph{{IEEE} Trans. Knowl. Data Eng.}, 32(8):1475--1488.

\bibitem[{Li et~al.(2022)Li, Yang, Wang, and Wei}]{DBLP:journals/corr/abs-2208-04232}
Zehan Li, Nan Yang, Liang Wang, and Furu Wei. 2022.
\newblock \href {https://doi.org/10.48550/arXiv.2208.04232} {Learning diverse document representations with deep query interactions for dense retrieval}.
\newblock \emph{CoRR}, abs/2208.04232.

\bibitem[{Liu et~al.(2020)Liu, Gong, Fu, Yan, Chen, Jiang, Lv, and Duan}]{DBLP:conf/acl/LiuGFYCJLD20}
Dayiheng Liu, Yeyun Gong, Jie Fu, Yu~Yan, Jiusheng Chen, Daxin Jiang, Jiancheng Lv, and Nan Duan. 2020.
\newblock \href {https://doi.org/10.18653/v1/2020.acl-main.604} {Rikinet: Reading wikipedia pages for natural question answering}.
\newblock In \emph{Proceedings of the 58th Annual Meeting of the Association for Computational Linguistics, {ACL} 2020, Online, July 5-10, 2020}, pages 6762--6771. Association for Computational Linguistics.

\bibitem[{Mehta et~al.(2022)Mehta, Gupta, Tay, Dehghani, Tran, Rao, Najork, Strubell, and Metzler}]{DBLP:journals/corr/abs-2212-09744}
Sanket~Vaibhav Mehta, Jai~Prakash Gupta, Yi~Tay, Mostafa Dehghani, Vinh~Q. Tran, Jinfeng Rao, Marc Najork, Emma Strubell, and Donald Metzler. 2022.
\newblock \href {https://doi.org/10.48550/arXiv.2212.09744} {{DSI++:} updating transformer memory with new documents}.
\newblock \emph{CoRR}, abs/2212.09744.

\bibitem[{Ren et~al.(2021)Ren, Qu, Liu, Zhao, She, Wu, Wang, and Wen}]{DBLP:conf/emnlp/RenQLZSWWW21}
Ruiyang Ren, Yingqi Qu, Jing Liu, Wayne~Xin Zhao, Qiaoqiao She, Hua Wu, Haifeng Wang, and Ji{-}Rong Wen. 2021.
\newblock \href {https://doi.org/10.18653/v1/2021.emnlp-main.224} {Rocketqav2: {A} joint training method for dense passage retrieval and passage re-ranking}.
\newblock In \emph{Proceedings of the 2021 Conference on Empirical Methods in Natural Language Processing, {EMNLP} 2021, Virtual Event / Punta Cana, Dominican Republic, 7-11 November, 2021}, pages 2825--2835. Association for Computational Linguistics.

\bibitem[{Robertson and Walker(1997)}]{DBLP:conf/sigir/RobertsonW97}
Stephen~E. Robertson and Steve Walker. 1997.
\newblock \href {https://doi.org/10.1145/258525.258529} {On relevance weights with little relevance information}.
\newblock In \emph{{SIGIR} '97: Proceedings of the 20th Annual International {ACM} {SIGIR} Conference on Research and Development in Information Retrieval, July 27-31, 1997, Philadelphia, PA, {USA}}, pages 16--24. {ACM}.

\bibitem[{Robertson and Zaragoza(2009)}]{DBLP:journals/ftir/RobertsonZ09}
Stephen~E. Robertson and Hugo Zaragoza. 2009.
\newblock \href {https://doi.org/10.1561/1500000019} {The probabilistic relevance framework: {BM25} and beyond}.
\newblock \emph{Found. Trends Inf. Retr.}, 3(4):333--389.

\bibitem[{Sun et~al.(2023)Sun, Yan, Chen, Wang, Zhu, Ren, Chen, Yin, de~Rijke, and Ren}]{DBLP:journals/corr/abs-2304-04171}
Weiwei Sun, Lingyong Yan, Zheng Chen, Shuaiqiang Wang, Haichao Zhu, Pengjie Ren, Zhumin Chen, Dawei Yin, Maarten de~Rijke, and Zhaochun Ren. 2023.
\newblock \href {https://doi.org/10.48550/arXiv.2304.04171} {Learning to tokenize for generative retrieval}.
\newblock \emph{CoRR}, abs/2304.04171.

\bibitem[{Sun et~al.(2020)Sun, Wang, Li, Feng, Tian, Wu, and Wang}]{DBLP:conf/aaai/SunWLFTWW20}
Yu~Sun, Shuohuan Wang, Yu{-}Kun Li, Shikun Feng, Hao Tian, Hua Wu, and Haifeng Wang. 2020.
\newblock \href {https://ojs.aaai.org/index.php/AAAI/article/view/6428} {{ERNIE} 2.0: {A} continual pre-training framework for language understanding}.
\newblock In \emph{The Thirty-Fourth {AAAI} Conference on Artificial Intelligence, {AAAI} 2020, The Thirty-Second Innovative Applications of Artificial Intelligence Conference, {IAAI} 2020, The Tenth {AAAI} Symposium on Educational Advances in Artificial Intelligence, {EAAI} 2020, New York, NY, USA, February 7-12, 2020}, pages 8968--8975. {AAAI} Press.

\bibitem[{Tay et~al.(2022)Tay, Tran, Dehghani, Ni, Bahri, Mehta, Qin, Hui, Zhao, Gupta, Schuster, Cohen, and Metzler}]{DBLP:conf/nips/Tay00NBM000GSCM22}
Yi~Tay, Vinh Tran, Mostafa Dehghani, Jianmo Ni, Dara Bahri, Harsh Mehta, Zhen Qin, Kai Hui, Zhe Zhao, Jai~Prakash Gupta, Tal Schuster, William~W. Cohen, and Donald Metzler. 2022.
\newblock \href {http://papers.nips.cc/paper\_files/paper/2022/hash/892840a6123b5ec99ebaab8be1530fba-Abstract-Conference.html} {Transformer memory as a differentiable search index}.
\newblock In \emph{NeurIPS}.

\bibitem[{Wang et~al.(2022)Wang, Hou, Wang, Miao, Wu, Chen, Xia, Chi, Zhao, Liu, Xie, Sun, Deng, Zhang, and Yang}]{DBLP:conf/nips/WangHWMWCXCZL0022}
Yujing Wang, Yingyan Hou, Haonan Wang, Ziming Miao, Shibin Wu, Qi~Chen, Yuqing Xia, Chengmin Chi, Guoshuai Zhao, Zheng Liu, Xing Xie, Hao Sun, Weiwei Deng, Qi~Zhang, and Mao Yang. 2022.
\newblock \href {http://papers.nips.cc/paper\_files/paper/2022/hash/a46156bd3579c3b268108ea6aca71d13-Abstract-Conference.html} {A neural corpus indexer for document retrieval}.
\newblock In \emph{NeurIPS}.

\bibitem[{Xiong et~al.(2021)Xiong, Xiong, Li, Tang, Liu, Bennett, Ahmed, and Overwijk}]{DBLP:conf/iclr/XiongXLTLBAO21}
Lee Xiong, Chenyan Xiong, Ye~Li, Kwok{-}Fung Tang, Jialin Liu, Paul~N. Bennett, Junaid Ahmed, and Arnold Overwijk. 2021.
\newblock \href {https://openreview.net/forum?id=zeFrfgyZln} {Approximate nearest neighbor negative contrastive learning for dense text retrieval}.
\newblock In \emph{9th International Conference on Learning Representations, {ICLR} 2021, Virtual Event, Austria, May 3-7, 2021}. OpenReview.net.

\bibitem[{Yang et~al.(2017)Yang, Fang, and Lin}]{DBLP:conf/sigir/Yang0L17}
Peilin Yang, Hui Fang, and Jimmy Lin. 2017.
\newblock \href {https://doi.org/10.1145/3077136.3080721} {Anserini: Enabling the use of lucene for information retrieval research}.
\newblock In \emph{Proceedings of the 40th International {ACM} {SIGIR} Conference on Research and Development in Information Retrieval, Shinjuku, Tokyo, Japan, August 7-11, 2017}, pages 1253--1256. {ACM}.

\bibitem[{Zhang et~al.(2022{\natexlab{a}})Zhang, Gong, Shen, Lv, Duan, and Chen}]{DBLP:conf/iclr/ZhangGS0DC22}
Hang Zhang, Yeyun Gong, Yelong Shen, Jiancheng Lv, Nan Duan, and Weizhu Chen. 2022{\natexlab{a}}.
\newblock \href {https://openreview.net/forum?id=MR7XubKUFB} {Adversarial retriever-ranker for dense text retrieval}.
\newblock In \emph{The Tenth International Conference on Learning Representations, {ICLR} 2022, Virtual Event, April 25-29, 2022}. OpenReview.net.

\bibitem[{Zhang et~al.(2022{\natexlab{b}})Zhang, Liang, Gong, Jiang, and Duan}]{DBLP:conf/acl/ZhangLGJD22}
Shunyu Zhang, Yaobo Liang, Ming Gong, Daxin Jiang, and Nan Duan. 2022{\natexlab{b}}.
\newblock \href {https://doi.org/10.18653/v1/2022.acl-long.414} {Multi-view document representation learning for open-domain dense retrieval}.
\newblock In \emph{Proceedings of the 60th Annual Meeting of the Association for Computational Linguistics (Volume 1: Long Papers), {ACL} 2022, Dublin, Ireland, May 22-27, 2022}, pages 5990--6000. Association for Computational Linguistics.

\bibitem[{Zhao et~al.(2022)Zhao, Liu, Ren, and Wen}]{DBLP:journals/corr/abs-2211-14876}
Wayne~Xin Zhao, Jing Liu, Ruiyang Ren, and Ji{-}Rong Wen. 2022.
\newblock \href {https://doi.org/10.48550/arXiv.2211.14876} {Dense text retrieval based on pretrained language models: {A} survey}.
\newblock \emph{CoRR}, abs/2211.14876.

\bibitem[{Zhou et~al.(2022)Zhou, Yao, Dou, Wu, Zhang, and Wen}]{DBLP:journals/corr/abs-2208-09257}
Yujia Zhou, Jing Yao, Zhicheng Dou, Ledell Wu, Peitian Zhang, and Ji{-}Rong Wen. 2022.
\newblock \href {https://doi.org/10.48550/arXiv.2208.09257} {Ultron: An ultimate retriever on corpus with a model-based indexer}.
\newblock \emph{CoRR}, abs/2208.09257.

\end{thebibliography}

\clearpage
\appendix

\section{Appendix}
\label{sec:appendix}

\subsection{Calculation method of error rate}
\label{sec:calculation of error rate}
Considering that the AR2 \citep{DBLP:conf/iclr/ZhangGS0DC22} itself does not make predictions on identifiers, we select identifier corresponding to the predicted document as AR2's identifier prediction. We calculate the error rate of model's prediction on the i$^{th}$ position as follows: For each predicting document identifiers, we calculate the probability that, given its prefix up to the i-$1^{th}$ position belonging to a prefix of a relevant document identifier, the addition of the model's prediction for the i$^{th}$ position no longer belongs to any prefix of a relevant document identifier.

\begin{table}[t]
    \renewcommand\arraystretch{1.2}
    \small
    \centering
    \begin{tabular}{lcc}
    \toprule[1pt]
    Dataset & Cluster Counts \\
    \hline
    NQ334K  & 34337 \\
    NQ1M  & 33003  \\ 
    NQ2M & 29000 \\
    NQ4M  & 28362 \\
    \bottomrule[1pt]
    \end{tabular}
    \caption{Total number of CIDs included in datasets with different scale of candidate document corpus.}
    \label{tab: magnitude of CIDs}
\end{table}

\subsection{Magnitude of CIDs}
\label{sec: Magnitude of CIDs}
We collect the total count of CIDs for datasets with different scales of candidate documents obtained through our proposed strategy introduced in section \ref{subsec:CID}. As shown in Table \ref{tab: magnitude of CIDs}, the results demonstrate that the proposed strategy can effectively control the total number of clusters, thus guarantee the memorizing volume of GDR. The reason why the magnitude of cluster counts in Table \ref{tab: magnitude of CIDs} (approximately 30000) is larger than the $Exp(|\mathrm{CID}|)$ we set as 5000 is that, The constructed cluster tree is unbalanced, resulting in more clusters than the expected value. Our preliminary studies show that, setting $Exp(|\mathrm{CID}|)$ in Algorithm \ref{alg:CID} as 5000 can lead to a favorable budget between efficiency and performance. 

\subsection{Experiments on TriviaQA}
We further verified the generalization of GDR on a subset of TriviaQA. We constructed the TriviaQA549K dataset following the procedure to construct NQ334K and compared GDR with other methods on it as shown in Table~\ref{tab:further}. The experimental results verified the good generalization of GDR.

\begin{table}[t]
    \renewcommand\arraystretch{1.2}
    \small
    \centering
    \begin{tabular}{lcc}
    \toprule[1pt]
    Method & Acc@20/100 & R@20/100 \\
    \hline
    BM25  & 95.7/98.6 & 70.5/93.2 \\
    AR2  & 97.9/99.3 & 71.8/92.8 \\ 
    NCI & 90.4/95.7 & 71.6/92.3 \\
    GDR  & 96.8/98.9 & 74.6/95.0 \\
    \bottomrule[1pt]
    \end{tabular}
    \caption{Experimental results on TriviaQA549K.}
    \label{tab:further}
\end{table}

\end{document}